\documentclass[12pt]{book}

\usepackage[dvips]{graphicx,color}
\usepackage{makeidx,tsukuba}

\makeauthorindex
\makeindex

\def\VEV#1{{\langle #1 \rangle}}
\def\lap{\;\rlap{\lower 2.5pt
   \hbox{$\sim$}}\raise 1.5pt\hbox{$<$}\;}
\begin{document}

\BookTitle{\itshape The 28th International Cosmic Ray Conference}
\CopyRight{\copyright 2003 by Universal Academy Press, Inc.}
\pagenumbering{arabic}

\chapter{
On the detectability of gamma-rays from Dark Matter annihilation in the Local Group with ground-based experiments
}

\author{%
%
%
Lidia Pieri$^{1,2}$ and Enzo Branchini$^1$ \\
{\it (1) Department of Physics, Roma Tre University, Rome, Italy \\
(2) INFN - Roma Tre, Rome, Italy} \\
}

\section*{Abstract}
Recent studies have suggested the possibility
that the lightest supersymmetric particle is a
suitable dark matter candidate. In this
theoretical framework, annihilations in high
density environments like the center of dark
matter haloes may produce an intense flux of gamma-rays.
In this paper we discuss the possibility of detecting the
signatures of neutralino annihilation in nearby galaxies with next
generation ground-based detectors.

\section{Introduction}

Revealing the nature of the Dark Matter (DM) is one
of the most challenging problems, from both a theoretical and experimental
viewpoint, facing particle physics and cosmology today.
The concordance model [8] requires that non-baryonic DM should contribute to
$\sim 26 \%$ of the universe content. Moreover, according 
to the Cold Dark Matter (CDM) paradigm, the bulk of DM
is required to be non-relativistic at the time of decoupling. 
A popular candidate for CDM is the Lightest Supersymmetric Particle (LSP). 
In most supersymmetry (SUSY)
breaking scenarios this is the neutralino $\chi$.
Assuming gaugino-universality, its mass is constrained by accelerator
searches and by theoretical considerations of thermal freeze-out
to lie in the range $50 \ GeV \lap m_{\chi} \lap 10 \ TeV$
[4,5].
If $R$-parity is conserved, the LSP can change its cosmological abundance
only through annihilation. \\
The Galactic Centre (GC) is the nearest high
density region and thus represents the most obvious site where to look for
DM annihiliation signals.
However, there are practical constraints: 
the GC is not visible at small zenith angles from sites
in the Northern hemisphere, where most of the experiments are located;
on the other hand, satellite instruments have small effective detection areas
for high energy photons, and can investigate only up to $\sim 300 \ GeV$. 
In this paper we study the sensitivity of ground-based experiments 
to $\gamma$-photon signals 
coming from the galaxies of the Local Group (LG).

\section{Signal and sources}

Expected photon fluxes from neutralino annihilation are given by 
\begin{eqnarray}
\frac{d \Phi_\gamma (E_\gamma)}{dE_\gamma}& = &
\left [ N_{\gamma \gamma} b_{\gamma \gamma} \delta (E_\gamma - m_\chi) +
N_{Z \gamma} b_{Z \gamma} \delta 
\left ( E_\gamma - m_\chi(1 - \frac{m_Z^2}{4 m_\chi^2}) \right ) + \right . 
\nonumber \\ 
& & \left . \sum_{F} \frac{d N_\gamma^F}{d E_\gamma} b_F \right ] 
 \frac{\VEV{\sigma v}_{a} }{m^2_\chi} \frac{1}{4 \pi D^2} \int_{0}^{r_{max}} \rho_{\chi}^2 (r) 
4 \pi r^2 dr.  
\label{flusso}
\end{eqnarray}
\noindent The first two terms in square brackets represent the $\gamma$-lines
with branching ratios of $\sim 10^{-3}$.
The photon flux is dominated by the continuum emission given
by the sum running over all the F final states.
At the tree-level neutralinos annihilate into fermions, gauge bosons, 
Higgs particles, gluons. Branching ratios depend on the assumed SUSY model.
Decay and/or hadronization in $\pi^0$ give a continuum spectrum of 
$\gamma$-photons emerging from the $\pi^0$ decay [2].
$\VEV{\sigma v}_{a}$ is the thermally-averaged 
annihilation cross-section.
The DM is assumed to be
concentrated in a single, spherical DM halo of radius $r_{max}$  
and density profile $\rho_\chi(r)$ located at distance $D$ from the observer.
The halo density profiles are poorly constrained by observations. 
In this work we adopted the Moore profile, which is supported by numerical
experiments [6]:
$\rho_\chi = \rho_s (\frac{r}{r_s})^{-1.5} (1 + (\frac{r}{r_s})^{1.5})^{-1}$,
where $\rho_s$ and $r_s$ are scale quantities;
it was truncated at the radius $r_{min}$ where 
the self-annihilation rate equals the dynamical time [1]. We imposed
$\rho_\chi(<r_{min}) = \rho_\chi(r_{min})$. \\
Since we are interested in nearby sources, we have considered
the nearest 45 LG galaxies, shown in the left panel of Fig.\,1.
The size of each circle is proportional to the expected
$\gamma$-ray flux within 1 squared degree, with a Moore density profile
and a cutoff radius $r_{min}$. Zenithal visibility from different sites 
on the earth of M31 and the GC is shown on the right panel.\\
\begin{figure}[t]
  \begin{center}
    \includegraphics[height=14pc,width=18.5pc]{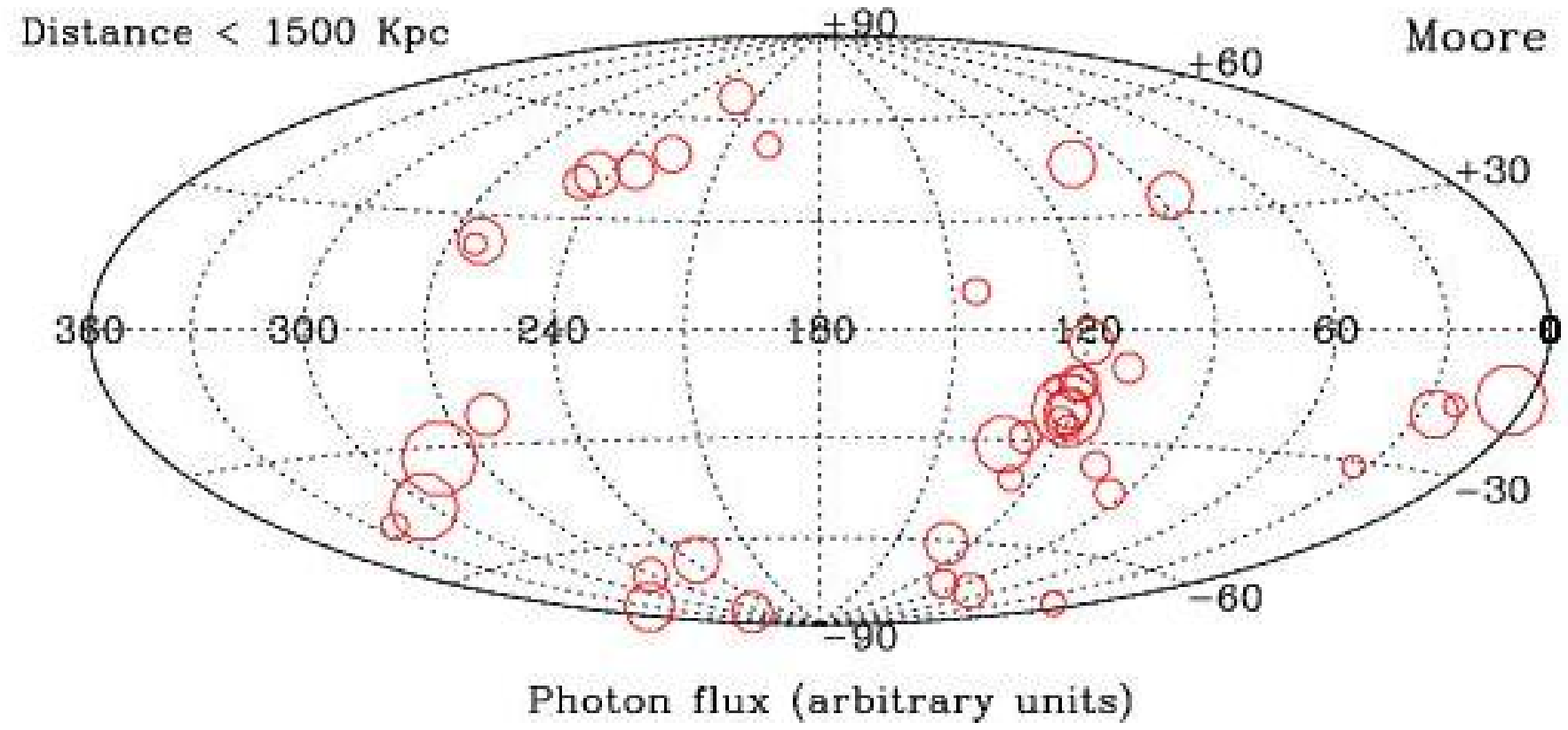}
    \includegraphics[height=14pc]{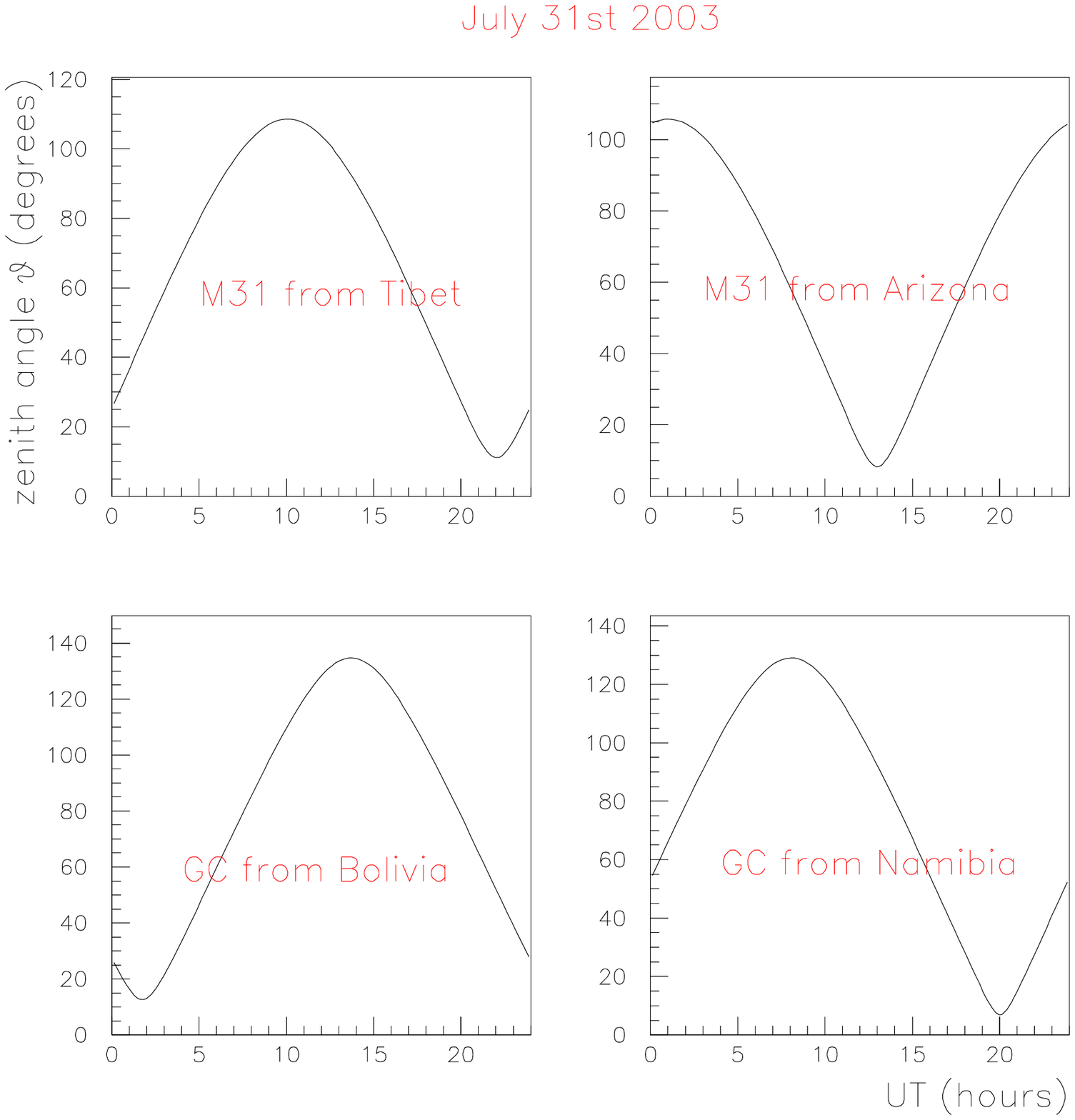}
  \end{center}
  \vspace{-1.5pc}
  \caption{Left: Aitoff projection of the 45 nearest LG galaxy distribution 
in Galactic coordinates.
The size of each circle is scaled to the $\gamma$-photons flux emitted 
within 1 squared degree from the halo center. 
Right: zenith angle under which M31
and the GC are seen from sites of various ground-based experiments.}
\end{figure}  
\noindent Hierarchical clustering in the CDM scenario predicts the presence of
sub-haloes that accrete into larger systems. 
DM haloes should host a population of sub-haloes with
a distribution function giving the probability of finding
a sub-halo of mass $m$ at a distance $r$ from the halo center:
$n_{sh}(r,m) = A (m / m_{H})^{-1.9} (1+\tilde{r}^2)^{-1.5}, \ \ 
\tilde{r} = r / r^{sh}_c $
where ${r^{sh}_c}$ is the core radius of the sub-haloes distribution,
$m_{H}$ is the mass of the parent halo and $A$ is a normalization constant [3].
\noindent Mass stripping and tidal heating modify both the size and the shape 
of sub-haloes. 
If we define the sub-halo tidal radius, $r_{tid}$, as the distance from 
the sub-halo center where
the tidal forces of the parent halo potential equal 
the self-gravity of the sub-halo.
One can assume that all the mass beyond $r_{tid}$ 
is lost in a single orbit without affecting its central 
density profile. 
In [7] the effect on the expected fluxes has been studied 
varying the mass clumped in sub-haloes, 
$m_{cl}$, the value of  ${r^{sh}_c}$, the minimum mass for sub-haloes, 
$m_{min}$, and the sub-haloes shape. Shallower density 
distributions and the presence of a black 
hole at the centre of the parent haloes have been taken into account as well. 
It is found that the fluxes from the Small and Large  
Magellanic Clouds (SMC, LMC) and M31 are well above the Galactic level for 
any choice of these parameters.
In some cases, like M33 and Sagittarius, extragalactic and Galactic 
contributions are comparable. 
In the left panel of Fig.\, 2 is shown the contribution to $\gamma$ fluxes 
from the MW smooth halo (dashed line) and LG galaxies above
the Galactic background (starred symbols) as a function of the
viewing angle $\psi$ from the GC.
The effect of including sub-haloes with tides calculated
at the circular orbit is also shown, both for the MW (solid line) and 
for the LG galaxies (big dots). $m_{cl} = 10\% \ m_H$, $r^{sh}_c = 30\%$ of 
the virial radius of the parent halo, $m_{min} = 10^6 M_\odot$, 
$m_\chi= 1 \ TeV$ and
$\VEV{\sigma v}_{a} = 10^{-25} \ cm^3 \ s^{-1}$ were assumed to obtain this
figure. 

\section{Ground-based detectability of DM photons}
The sensitivity of experimental apparates is computed by comparing 
the number of $\gamma$ events expected from the source to the fluctuations
of background events. 
Due to the low signal level, 
the electron and 
diffuse $\gamma$-ray backgrounds must be taken into account 
besides the usual hadron background:
\begin{equation}
\frac{n_{\gamma}}{\sqrt{n_{bkg}}}= \frac{\sqrt{T_\delta} \epsilon_{\Delta \Omega} \int \int \epsilon_\gamma A^{eff}_\gamma (E,\theta) \frac{d\phi^{DM}_\gamma}{dE} dE d\theta}{\sqrt{ \Delta \Omega \int \int \left [ (1 - \epsilon_{h}) A^{eff}_{h} (E,\theta) \frac{d\phi_{h}}{dE} + \epsilon_\gamma A^{eff}_\gamma (E,\theta) \left (\frac{d\phi_{e}}{dE} + \frac{d\phi_{\gamma}^{diff}}{dE} \right) \right ] dE d\theta }}.
\end{equation}
$T_\delta$ is the time during which the source is seen with zenith angle
$\theta \leq 30^o$, 
$\epsilon_{\Delta \Omega} = 0.7$ is the fraction of signal events 
within the optimal solid angle $\Delta \Omega$, $\epsilon_{\gamma,h}$ 
are the identification efficiencies of showers induced by photons 
or hadrons. 
Sensitivities for a $\sim 80 \times \ 80 \ m^2$ high altitude full
coverage array (effective areas $A^{eff}_{\gamma,h}$ were taken from [9]) 
and for a high altitude Cherenkov cell (an effective area of
$10^4 \ m^2$ has been assumed) have been computed.
The resulting curves are plotted in the right panel of Fig.\, 2 (solid lines), 
where expected $\gamma$ flux from $\chi \chi$ annihilation is also shown for 
comparison (dot-dashed curve). \\
\begin{figure}[t]
  \begin{center}
    \includegraphics[height=15.5pc]{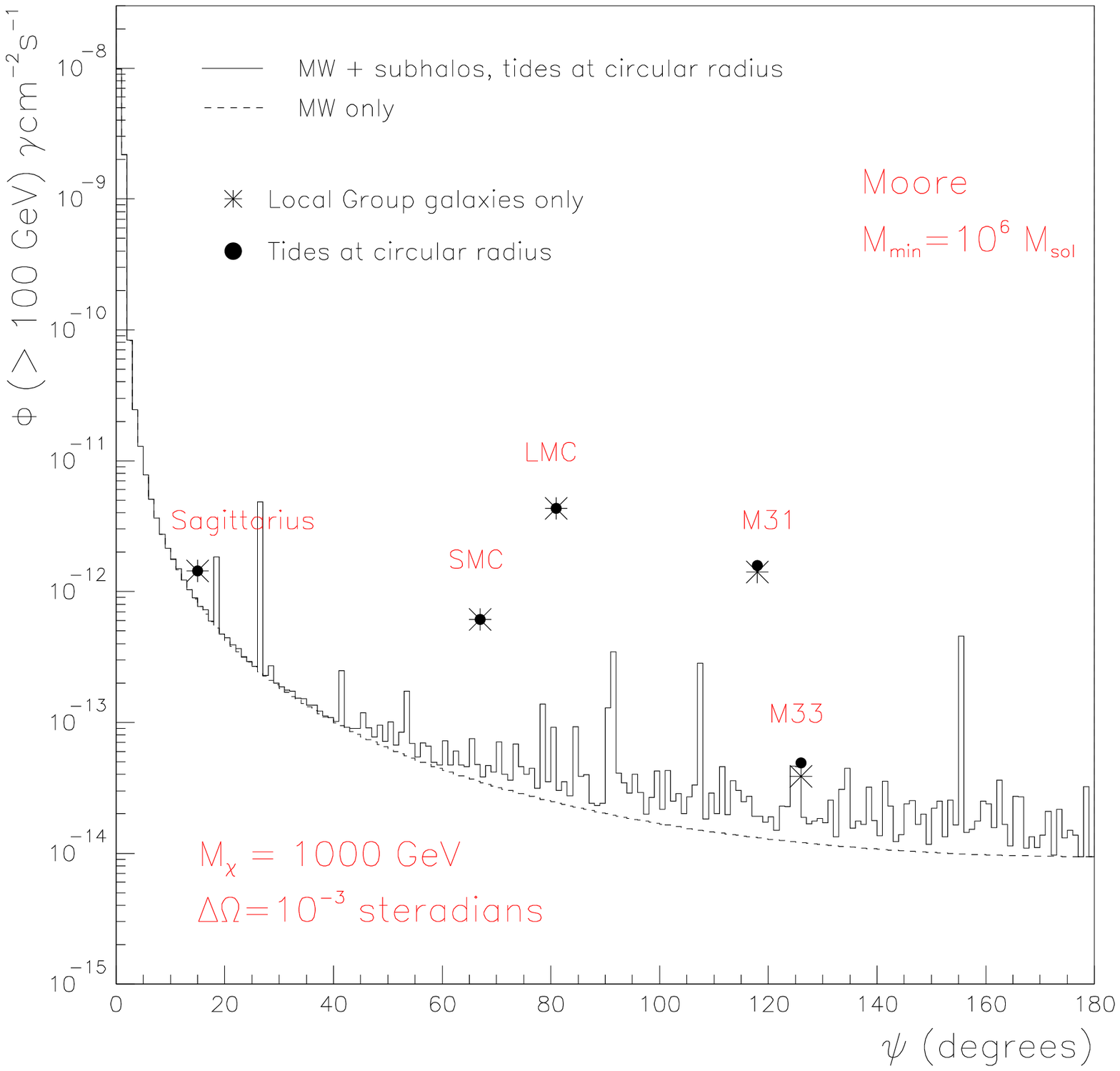}
    \includegraphics[height=15.5pc]{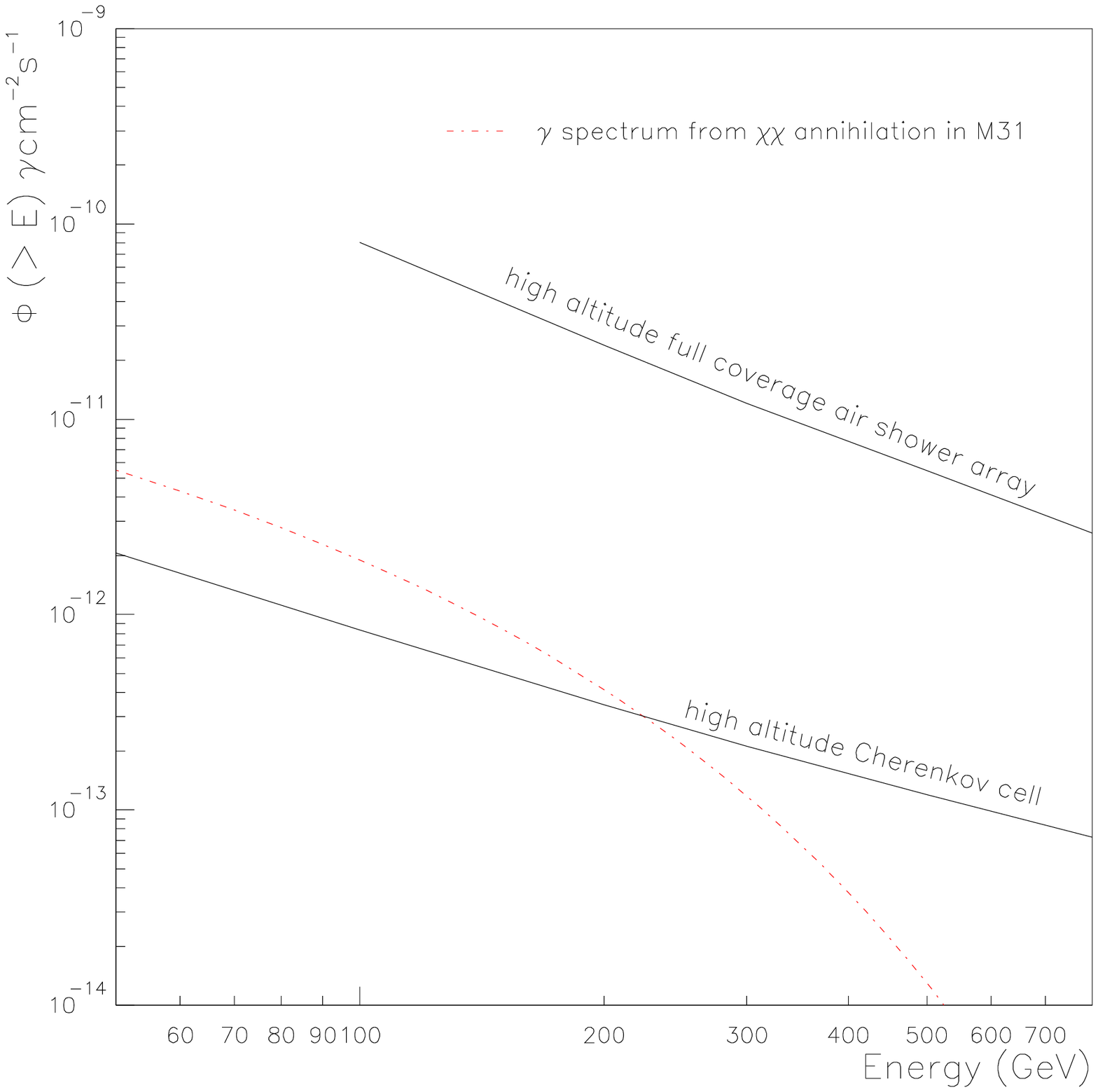}
  \end{center}
  \vspace{-1.5pc}
  \caption{Left: integrated $\gamma$-photon flux 
$> 100 \ GeV$ from neutralino annihilation.
Right: $5 \sigma$ sensitivity of a high altitude full
coverage air shower detector array (in 1 year of data taking, $\epsilon_{\gamma,h} = 75 \%$, $\Delta \Omega = 10^{-3} \ sr$) 
and of a high altitude 
Cherenkov cell (in 20 days, $\epsilon_{\gamma,h} = 99 \%$, $\Delta \Omega = 10^{-5} \ sr$) to
$\gamma$-photons from $\chi \chi$ annihilation in M31.}
\end{figure}

In conclusion, we found that DM annihilation signatures in extragalactic
dense objects could be revealed with a significance of 5 $\sigma$ with
next generation ground-based detectors mainly based on the Cherenkov 
tecnique. \\

\noindent {\sl Acknowledgements.}
We thank M. De Vincenzi and T. Di Girolamo for useful comments and suggestions.

\vspace{\baselineskip}

\re
1.\  Berezinsky V., et al., Phys. Lett. B 294, 221 (1992).
\re
2.\  Bergstrom L., et al., Nucl. Phys. Proc. Suppl. 81, 22, (2000).
\re
3.\  Blasi P. and Sheth R. K., Phys. Lett. B 486, 233 (2000).
\re
4.\  Ellis J., et al., Phys. Rev. D. 62, 075010 (2000).
\re
5.\  Jungman G., et al., Phys. Rep. 267, 195 (1996).
\re
6.\  Moore B., et al., Astrophys. J. Lett. 524, L19 (1999).
\re
7.\  Pieri L. and Branchini E., 2003.
\re
8.\  Spergel D. N., et al., astro-ph/0302209.
\re
9.\  Vernetto S. and Turasco L., ARGO internal note ARGO-YBJ/2003-001. 
\endofpaper
\end{document}